# Engineering bandgaps of monolayer MoS$_2$ and WS$_2$ on fluoropolymer substrates by electrostatically tuned many-body effects

**Bo Liu[1,2†], Weijie Zhao[2,3†], Zijing Ding[1,4], Ivan Verzhbitskiy[3], Linjun Li[1,2], Junpeng Lu[3], Jianyi Chen[1], Goki Eda[1,2,3], Kian Ping Loh[1,2*]**

[1]Department of Chemistry, National University of Singapore, 3 Science Drive 3, 117543 Singapore

[2]Centre for Advanced 2D Materials and Graphene Research Centre, National University of Singapore, 6 Science Drive 2, 117546 Singapore

[3]Department of Physics, National University of Singapore, 2 Science Drive 3, 117542 Singapore

[4]SZU-NUS Collaborative Innovation Center for Optoelectronic Science and Technology, and Key Laboratory of Optoelectronic Devices and Systems of Ministry of Education and Guangdong Province, College of Optoelectronic Engineering, Shenzhen University, Shenzhen 518060, People's Republic of China

[†]These authors contributed equally to this work.

[*]Correspondence to: chmlohkp@nus.edu.sg

## Keywords:

Monolayer group VI transition metal dichalcogenides (TMDCs), denoted as $MX_2$ where M = Mo, W; X = S, Se, have emerged as a new class of two-dimensional (2D) semiconductors and attracted extensive research interest after graphene due to their sizeable band-gap and remarkable optical and electronic properties[1-9]. The significantly suppressed dielectric screening at the limit of monolayer thickness gives rise to tightly bound excitons with exceptionally large binding energy ($E_b$))[10-12, 13-16]. The tightly bound excitons, as well as associated trions (charged excitons)[4, 5] and biexcitons[6, 17] render TMDC an ideal platform to investigate fundamental many-body physics, and its optical and electrical properties can be engineered for potential applications in nanoelectronics[7, 18, 19], photonics[3, 6, 8, 13] optoelectronics[20] and valleytronics[21-24].

The optical and excitonic properties of monolayer $MX_2$ are known to be highly susceptible to the substrate and carrier doping[9, 10, 15, 25-27, 28-30] due to their 2D nature. For example, substrate-induced dielectric screening effect can strongly renormalize the quarsi-particle bandgap ($E_g$) and reduce the exciton binding energy of monolayer $MX_2$ through modifying electron-electron and electron-hole Coulomb interactions[10, 11, 15, 26, 28, 31]. The significant bandgap renormalization effect (BGR) and the reduction of exciton binding energy induce opposite shifts of optical bandgaps and cancel each other mostly[10, 26, 28, 31]. Therefore, only slight shifts of optical bandgaps (the so-called A exciton) were observed by varying dielectric environment[26, 28, 31]. In addition, substrates could also introduce inhomogeneous doping to 2D materials through charge transfer or trapped impurities[30, 32, 33, 34].

On the other hand, carrier doping provides a dynamic control on optical properties of monolayer $MX_2$ [4-6, 9, 21, 29]. With the increase of electron doping concentration, as an example, free electrons start to fill the lowest conduction band of monolayer $MX_2$ at the K/K' points at the Brillouin zone, therefore several many-body effects are essentially involved. First of all, phase-space filling effect arising from Pauli exclusion principle because of the fermionic nature of electrons and holes would gradually drive the onset of direct dipole transition to higher energies (referred as Pauli blocking effect in the following discussion, $\Delta E_{PB}$)[4, 35, 36], reduce the exciton oscillator strength and further lower the exciton binding energy ($\Delta E_b(PSF)$)[9, 35, 36]. Secondly, doping-induced free carrier plasmons are able to couple with quarsi-particles, resulting in the electronic bandgap renormalization ($\Delta E_g(BGR)$) and reduction of exciton binding energy($\Delta E_b(CP)$) through an effective screening effect[9, 29, 35, 36]. Thirdly, with the presence of free electrons, exciton will capture an extra electron to form a three-body quarsi-particle, i.e. trion, then the oscillator strength of excitons and PL spectra weight transfers to trions progressively[4, 5, 21, 25, 26, 37]. The interplays between these many-body effects determine the optical and excitonic properties of monolayer $MX_2$ as a function of carrier doping density. However, unlike the dielectric screening effects from substrates that have been well understood both theoretically and experimentally[10, 11, 26, 28], there is limited work[4, 9] investigating carrier doping effect of monolayer $MX_2$ systematically, owing to the intricate interplays between associated many-body effects. Some essential problems, such as: the impact of phase-space filling effect, the influence of dark states[9, 16, 38, 39] and defect states in monolayer $MX_2$ and intrinsic optical bandgap evolution as a function of doping carrier density, are not well understood.

Here we systematically studied the carrier doping effect in monolayer $MoS_2$ and $WS_2$ using back-gated field-effect transistor (FET) devices by taking advantages of the fluoropolymer CYTOP[TM] as a high quality substrate (see detailed discussion on CYTOP in the supporting information). Differential reflectance and micro photoluminescence (PL) spectroscopy were used to monitor the evolution of optical bandgaps and exciton/trion dynamics as a function of carrier doping concentration. For electrostatically gated monolayer $MoS_2$ and $WS_2$, giant bandgap renormalization dominates the reduction of exciton binding energy and leads to a red-shift in exciton energies with increasing carrier doping. The Pauli blocking effect can cause a blue-shift of optical bandgap effectively in electron doped monolayer $MoS_2$. While in monolayer $WS_2$, the robust bandgap renormalization dominates the other many-body effects and subsequently results in a red-shift of optical bandgap. Furthermore, a high density of defect states can effectively reduce the carrier doping effect by trapping free carriers and facilitate the formation of much "heavier" negative trion with exceptionally large binding energy in electron doped monolayer $WS_2$.

Thin $MX_2$ flakes were exfoliated from bulk crystals using blue tapes and then deposited on a polydimethylsiloxane (PDMS) viscoelastic stamp[40] (instead of silicon wafer). The monolayer flakes are identified on PDMS by optical contrast using an optical microscope and confirmed by micro PL and Raman measurements. The flakes can then be easily aligned and transferred onto various substrates using PDMS as the transfer media. Mild heating facilitates the transfer, and the PDMS is released from the target substrate very slowly to minimize the flake deformation. When deposited onto various substrates, monolayer $MX_2$ typically exhibits inhomogeneous doping unintentionally and dielectric screening effect from substrates[26, 32, 33]. In order to obtain high quality substrates, we tested a wide range of substrates and checked PL spectra of monolayer $MoS_2$ transferred on them, as shown in **Figure 1a** and Figure S1 in the supporting information.

The tested substrates include $SiO_2$, $Al_2O_3$, poly(methyl methacrylate) (PMMA), poly(vinyl alcohol) (PVA), h-BN (annealed), PDMS, CYTOP/$SiO_2$, and a suspended $MoS_2$ sample was used for comparison. According to the PL spectra shown in Figure 1a, a narrow and strong PL peak at the region of A exciton, without the trion peak present, was obtained for monolayer $MoS_2$ on the CYTOP/$SiO_2$ substrate. This suggests CYTOP is an ideal substrate for 2D crystals, on account of its very low surface energy, low surface trap densities, absence of surface dangling bonds as well as low permittivity ($\varepsilon$= 2.1), which will minimize substrate dielectric screening effect (see detailed analysis in supporting information).

The CYTOP substrate in this work is a bilayer stack of a 20 nm thick CYTOP (chemical structure in Figure S2) film spun cast on 300 nm silicon oxide/p+ silicon substrates (see Experimental Section). The typical optical image and corresponding atomic force microscope (AFM) image of monolayer $MoS_2$ on $CYTOP/SiO_2/Si$ are shown in Figure 1b and the inset, respectively, and the AFM image of CYTOP surface is shown in Figure S2. The CYTOP polymer presents very flat surface with surface roughness (RMS) of about ~0.5nm, comparable to silicon oxide surface, which is an important factor for obtaining high quality TMDC flakes without causing ruptures, tensions or bubbles. Moreover, we found that the adhesion of $MX_2$ flakes to the CYTOP layer is strong enough for reliable optical measurements and device fabrication. In the following, we characterized the optical and electrical properties of monolayer $MoS_2$ and $WS_2$ on the CYTOP substrate and further investigated their intriguing carrier doping induced many-body physics.

In **Figure 2a** and 2b, the excitonic features of monoayer $MoS_2$ and $WS_2$ on quartz (i.e. $SiO_2$) and CYTOP/quartz were investigated by differential reflectance and PL measurements, respectively. Differential reflectance is defined as $\frac{\Delta R}{R} = 1 - \frac{R_1(\lambda)}{R_0(\lambda)}$, where $R_1(\lambda)$ and $R_0(\lambda)$ are the reflectance of TMDCs flakes on the substrate and the substrate itself, respectively. The differential reflectance can be used to derive the absorption spectrum for very thin films[41] such as monolayer to few-layer $MX_2$[3, 42]. Monolayer samples are mechanically exfoliated on quartz substrates for reflectance measurements without any further processing. The thickness of the CYTOP layer is intentionally reduced to 20nm to suppress the interference effect in the reflectance spectrum. In Figure 2a and 2b, the excitonic features (A and B excitons) of monolayer $MoS_2$ and $WS_2$ are well captured by differential reflectance and PL spectra. We note that the Stokes shift (i.e. the energy difference between absorption and emission peaks) of A excitons (corresponding to its optical bandgap) is around 25 and 15 meV for $MoS_2$ and $WS_2$ samples, respectively, which are consistent with previous literature reports [43]. Strikingly, monolayer $MoS_2$ and $WS_2$ on CYTOP/quartz substrates show negligible Stokes shift, as shown in Figure 2b. The full width at half maximum (FWHM) of the PL spectrum of $MoS_2$ and $WS_2$ on CYTOP are 45 and 27 meV respectively, which are much narrower than that of $MoS_2$ (80 meV) and $WS_2$ (50 meV) samples on $SiO_2$/Si or quartz. These results imply the large Stokes shifts of monolayer $MoS_2$ and $WS_2$ on silicon dioxide substrates mainly arise from the spectra weight reduction and broadening of PL peaks caused by the inhomogeneity of substrate-induced doping[5, 32, 33, 34], rather than the electro-static doping itself[4]. The inhomogeneous doping leads to spatial fluctuation of exciton and trion states, as depicted schematically in the inset of Figure 2a. Consequently, PL originates at the lowest optical band-gap region as the electrons and holes tend to relax to the lowest conduction and valence states, respectively, in a shorter time scale than the radiative recombination lifetime [4, 5, 44]. On the contrary, the absorption peak reflects the integrative optical density of states. The FWHM of $MoS_2$ on CYTOP is still higher than that of $WS_2$ on CYTOP, the broadening is probably due to a relatively higher intrinsic doping or defect density[4, 45].

We now turn to the charge transport properties of monolayer $MoS_2$ and $WS_2$ FET devices fabricated on CYTOP/$SiO_2$/Si substrates. FET devices on $SiO_2$/Si were demonstrated for comparison. The hydrophobicity of CYTOP surface makes it hard to apply conventional photolithography or e-beam lithography to deposit electrical contacts on the 2D flakes. Surface treatment is ruled out since it changes the interaction between treated CYTOP and the flakes. Here we use micro stencil masks (See supporting information Note 4 and Figure S4 for fabrication details, microscope image of metal contacts on a typical device, respectively) to directly evaporate 50 nm gold onto the flakes for electrical contacts.

A typical FET device of monolayer $MoS_2$ on CYTOP/$SiO_2$/Si shows n-type trans-conductance and threshold voltage of ~-7 V by tuning silicon back gate ($V_{bg}$), as shown in Figure 2c. P-type conductivity is not clearly observed, although the source-drain current increases at around -60 V back gate voltage, probably due to the high Schottky barrier between $MoS_2$ valence band (<-5.8 eV[46]) and the gold contact Fermi level (-5.0 eV). The field effect mobility is estimated from the linear region of the $I_d$ - $V_g$ curve to be 20 cm$^2$/Vs with a two-terminal configuration using the following equation:

$$\mu = \frac{L}{W} \cdot \frac{1}{CV_d} \cdot \frac{dI_d}{dV_g},$$

where L and W are channel length and width, respectively, C is the gate capacitance of 300 nm SiO$_2$ or 300 nm SiO$_2$ with 20 nm CYTOP (equivalent to ~340 nm SiO$_2$), $V_d$ is the drain bias, $I_d$ is the drain current, and $V_g$ is the gate voltage. The MoS$_2$ on SiO$_2$/Si FET devices were fabricated with the same shadow mask method but without CYTOP layer, thus the possible contamination of photo-resist or e-beam resist is eliminated. The $I_d$-$V_g$ curve shows much heavier n- doped characteristics and negatively shifted threshold voltage of ~-40 V and mobility of ~10 cm$^2$/Vs, consistent with previous reports[7, 18].

In the case of monolayer WS$_2$, the $I_d$-$V_g$ curves of FET devices on CYTOP and on SiO$_2$/Si are compared in Figure 2d. The device on SiO$_2$/Si shows heavily n-doped characteristics and the field effect mobility was estimated to be 12 cm$^2$/Vs. On the contrary, the device on CYTOP exhibits weaker n-type conductivity and ambipolar behavior with electron and hole mobility of 15 and 3 cm$^2$/Vs, respectively. We note the carrier mobility can vary in the range of 1~40 cm$^2$/Vs and the threshold voltages have a variation of ±5V for different devices fabricated. N-type characteristics in monolayer WS$_2$ FET has been extensively reported [9, 19] but the ambipolar characteristics has only been observed with ion gel gating with ultrahigh doping densities[47]. To the best of our knowledge, we have demonstrated ambipolar characteristics in exfoliated monolayer $WS_2$, for the first time, with conventional solid state gating on CYTOP substrates without intentional doping, which can facilitate studies of many-body physics and novel valley contrasting phenomena, such as valley hall effect [22, 23] and valley pseudospin[21, 23, 24], and has great potential to be implemented on a large scale.

The n-type characteristics in monolayer $MoS_2$ and $WS_2$ are mainly attributed to substrate induced doping and intrinsic chalcogenide vacancies [32, 48] (detailed analysis in supporting information). The former is due to charge transfer from $SiO_2$ substrate and trapped impurities at the interface [32, 33, 34], which can be considered as the dominant doping source of monolayer 2D crystals on $SiO_2$ substrates. CYTOP can be used as a high quality substrate for FET fabrication as it has significantly minimized substrate induced doping and also provides weak electrostatic screening due to its low dielectric constant. Here, we investigate the intrinsic carrier doping effects of monolayer $MoS_2$ and $WS_2$ on CYTOP/$SiO_2$/Si substrates through electrostatic gating.

We first examine monolayer $MoS_2$ FET devices on CYTOP substrates, since it is the most intensively studied material in the $MX_2$ family both experimentally and theoretically. **Figure 3a** and inset show the gate tunable PL spectrum of $MoS_2$ FET device on both 20 nm CYTOP and 300 nm $SiO_2$ dielectric layers with heavily doped silicon back gate at room temperature. Consistent with the electrical transport measurements, the neutral point of monolayer $MoS_2$ on CYTOP is largely shifted to lower negative voltages due to the suppression of substrate induced electron doping, allowing the tunability of PL intensity to be greatly improved. The PL intensity of A excitons, as well as electron concentration in FET devices shown in Figure 2c, experiences a sudden change of $V_{bg}$ from -20 V to 0 V for $MoS_2$ on CYTOP. In contrast to CYTOP, the $V_{bg}$ tunable PL is also observed in $MoS_2$ sample on $SiO_2$/Si, but with a lower slope from -80V to -60 V, suggesting that the charge neutrality point is shifted by more than 50 V, or charge concentration of ~$4x10^{12}$ $cm^{-2}$. We also note that the PL spectrum at $V_{bg}$ = 40~60 V resembles the PL spectrum of $MoS_2$ on quartz or $SiO_2$ substrates in Figure 1a, indicating the extrinsic doping level caused by quartz or $SiO_2$ can be estimated to be $3~5x10^{12}$ $cm^{-2}$. However, this value may be dependent on the processing conditions of the silicon or quartz substrates.

To resolve the excitonic features of monolayer $MoS_2$ on CYTOP, we measured low temperature (77K) differential reflectance as shown in Figure 3b. PL (shown in Figure S9, supporting information) was also recorded as a function of back gate voltage. Based on the FET transfer curve in Figure 2c, the gate dependent PL in Figure 3a and Figure S9 and reflectance peak area analysis in Figure S10, the charge neutrality point of $MoS_2$ on CYTOP can be estimated to be -40 V, which allows the electron doping concentration to be estimated by using ne=$CV_{bg}$ [4]. The extracted peak positions of A excitons and negative trions in gate tunable reflectance spectra (see details in supporting information) are plotted in Figure 3c against $V_{bg}$. The A exciton blue-shifts at relative low doping, then start to saturate when electron doping exceeds $3.8x10^{12}$ $cm^{-2}$, while the trion peak red-shifts generally. As discussed above, for neutral $MoS_2$, the A exciton energy (referred as optical bandgap, $E_{opt}$) is determined by:

$$E_{opt} = E_g - E_b,$$

While at finite electron doping, the change in the $E_{opt}$ can be expressed as:

$$\Delta E_{opt} = \Delta E_g(BGR) + \Delta E_{PB} - \Delta E_b, \text{ with}$$

$$\Delta E_b = \Delta E_b(PSF) + \Delta E_b(CP),$$

where $\Delta E_g(BGR)$ is the carrier doping induced quarsi-particle bandgap renormalization, $\Delta E_b(PSF)$ and $\Delta E_b(CP)$ are the change of exciton binding energy owing to phase-space filling effect and carrier screening effect, respectively, and $\Delta E_{PB}$ is the change of direct dipole transition onset due to Pauli blocking[4] :

$$\Delta E_{PB} = \hbar^2 \pi n / 2\mu e^2,$$

where $\hbar$ is Planck constant, n is the electron density, e is the elementary charge and $\mu$ is exciton reduced mass: $1/\mu = 1/m_e + 1/m_h$, $m_e$ and $m_h$ are the effective electron and hole masses at the band edge near K/K' point, respectively with $m_e$ = $0.5*m_0$ and $m_h$ = $0.6*m_0$ [44], where $m_0$ is the electron rest mass. The calculated $\Delta E_{PB}$ at relative low doping density (below 3.0 x$10^{12}$ cm$^{-2}$ or $V_{bg}$ < 45 V ) were plotted in Figure 3c, since the conduction band can be treated as a parabolic dispersion within very low momenta close to K/K' points[16, 44] and dark exciton states seats slightly above the bright A exciton states[16, 23, 44, 49] as schematically demonstrated in Figure 3d. As shown explicitly in Figure 3c, the $\Delta E_{PB}$ is responsible for the blue-shift of $E_{opt}$ by overwhelming the red-shift of $E_{opt}$ arising from the incomplete compensation of the bandgap renormalization and reduction of exciton binding energy. However, when electron doping exceeds $3.0 \times 10^{12}$ cm$^{-2}$, the non-parabolicity of the exciton dispersion curve and the filling of dark exciton states start to be effective and strongly reduce the uprising amplitude of $\Delta E_{PB}$, therefore the saturation and following drop of $E_{opt}$ is observed.

In order to evaluate the impact of bandgap renormalization effect against electron doping density, we extract the changing rate of $\Delta E_b$ against doping density from Zhang's recent theoretical work [44] on monolayer $MoS_2$ by taking both phase-space filling and carrier screening effects into account, which reliably reproduces experimental results and demonstrates the changing rate of $\Delta E_b$ is not affected by effective dielectric constant generally. Since the extracted changing rate of $\Delta E_b$ is -24 meV/$10^{12}$ cm$^{-2}$[44], combined with that of $\Delta E_{PB}$ being 4.3 meV/$10^{12}$ cm$^{-2}$ and $\Delta E_{opt}$ being 3.0 meV/$10^{12}$ cm$^{-2}$ shown in Figure 3c, the changing rate of $\Delta E_g(BGR)$ can be calculated to be -25.3 meV/$10^{12}$ cm$^{-2}$ for electron doping less than $3.0 \times 10^{12}$ cm$^{-2}$. Consequently, the quarsi-particle bandgap of monolayer $MoS_2$ is effectively renormalized in a large magnitude of -114 meV with an electron doping density of $5 \times 10^{12}$ cm$^{-2}$, as a result of highly suppressed dielectric screening in the 2D $MX_2$ system. Furthermore, the magnitude of $\Delta E_b(PSF)$ is estimated based on a theoretical model[35] of transferring the phase-space filling effect into an effective dielectric constant, resulting in:

$$\Delta E_b(PSF) = E_b[1 - \left(1 - \frac{n}{2n_c}\right)^2],$$

$$\approx E_b * n/n_c \text{ , when } n_c >> n,$$

where, n is the electron doping concentration and $n_c$ is the critical doping density at which the exciton is completely quenched by the phase-space filling effect and is determined by exciton Bohr radius. The relative small Bohr radius of A exciton, which is about 9.3 Å[14, 44] signifies the exciton wave function spreads to large momenta, suggesting it can sustain over very high doping density at the point of view of the phase-space filling effect. Here, $n_c$ is calculated to be $7.4 \times 10^{13}$ cm$^{-2}$ using

$$n_c = 2/\pi a^2,$$

The changing rate of $\Delta E_b$(PSF) is obtained as ~-6.8 meV/$10^{12}$ cm$^{-2}$, smaller than that of $\Delta E_b$(CP) induced by carrier screening (-17.2 meV/$10^{12}$ cm$^{-2}$). The prominent tunability of the fundamental bandgaps and exciton binding energy makes MX$_2$ promising candidates for novel optoelectronic applications.

Due to the limited access of hole doping region and broad exciton/ trion peaks, the evolution of exciton and positive trion states cannot be resolved unambiguously for monolayer MoS$_2$. Therefore, we turn to monolayer WS$_2$ with well implemented ambipolar transport behavior and narrow exciton/ trion peaks.

**Figure 4a** shows the evolution of differential reflectance spectra of monolayer WS$_2$ as $V_{bg}$ is scanned from +70 V to -70 V and the A exciton and trion peaks can be well resolved and their peak positions are summarized in Figure 4b. The optical bandgap of monolayer WS$_2$ red-shifts at both electron and hole doping regime, with a changing rate of -1.1 meV/$10^{12}$ cm$^{-2}$ and -2.7 meV/$10^{12}$ cm$^{-2}$, respectively, suggesting that the bandgap renormalization effect is quite robust and dominates the other many-body effects. However, at the electron doping regime, we might expect a higher rate of optical bandgap red-shift because of the presence of dark exciton states below the lowest bright A exciton states[16], as shown in the schematic diagram in Figure 4c. It is well established that the strong spin-orbital coupling in monolayer $MX_2$ gives rise to large valence band splitting and the associated A and B excitons at around K/K' points of the Brillouin zone[2-4, 14, 49, 50]. On the contrary, the conduction band splitting is usually overlooked due to its very small values, i.e. about 5 meV for monolayer $MoS_2$ and 30 meV for monolayer $WS_2$,[16, 39] respectively.

In monolayer $WS_2$, optical transitions between the lowest conduction and top valence bands are prohibited due to their anti-parallel spin polarizations, resulting in the so-called dark excitons[16, 38, 39] which significantly affects its optical and electrical properties[9, 16, 38, 39]. Here, at relative low electron doping concentration, electrons mostly fill the dark exciton states, thus Pauli blocking and phase-space filling causes little effects for bright A excitons. Subsequently one would expect a higher changing rate of A exciton red-shift against electron doping density. The low changing rate observed in this work indicates that the carrier doping effect has been weakened, probably due to the intrinsic defects in $WS_2$. Indeed, we have observed strong PL from "localized states" of defects and impurities in Figure S11 and negative trions with abnormally large binding energy shown in

Figure 4b. The localized states in $MX_2$, like chalcogen vacancies, play a key role in the free carrier and exciton/trion dynamics and the optical transitions[4, 5, 51, 52]. When the electrons are injected into $WS_2$ electrically, these defect states are able to efficiently capture free electrons in an ultrafast process (within several pico-seconds)[51], making electrons less effective in the carrier doping induced many-body effects[53, 54] and potentially affecting the formation of negative trions[55]. In Figure 4b, the large difference between the positive ($E_{A+}$, 21 meV) and negative trion binding energies ($E_{A-}$, 43 meV), estimated from the energy differences between neutral exciton and trion peaks at around zero doping[4, 54], cannot be attributed to the slight difference between electron and hole effective masses in the mass action model[5, 56, 57]. As estimated here, for monolayer $WS_2$, the ~20% difference in electron and hole effective masses[16, 49, 50] can only account for less than 10% difference in trion binding energy. The negative trion consists of one electron in the conduction band, another electron in the LS and one hole on the valence band, as shown in the inset of Figure 4b schematic, thus it is much heavier than the trion involving only electrons in the conduction band [53, 54]. The electron on the localized states can be treated as immobile negative charge and the Hamiltonian of the negative trion can be modified accordingly[37, 55, 56]. In a simplified presentation, the negative trion binding energy can be written as the following in the presence of localized states[55, 56]:

$$\frac{E_{A^-}}{D_N} = \frac{1}{1+\sigma}\left[\left(\frac{3}{2}\right)^2\left(\frac{\sigma}{\sigma^2+3\sigma+1}+1\right)^{-1}-1\right]$$

where $D_N = m_e R_H / \varepsilon^2 m_0$, $\sigma = m_e / m_h$. $m_e$ and $m_h$ are electron and hole effective mass, $R_H$ is the Rydberg constant, $\varepsilon$ is the dielectric constant and $m_0$ is the electron rest mass. In the absence of localized states[55, 56],

$$\frac{E_{A^-}}{D_N} = \frac{1}{1+\sigma}\left[\left(\frac{3}{2}\right)^2\left(\frac{2\sigma+1}{\sigma^2+4\sigma+2}+1\right)^{-1}-1\right]$$

For monolayer $WS_2$, $\sigma$ is around 0.8[16, 37, 50], and $E_{A^-}$(with localized states)/$E_{A^-}$ = 1.58, in agreement with our experimental results qualitatively. Accurate quantitative analysis requires other details to be considered, such as the defect types and intensities, and more complex theoretical models are needed [37, 55, 56].

In conclusion, our results suggest that CYTOP can serve as an ideal platform to study intrinsic electrical and optical properties in TMDCs and other low dimensional systems that are susceptible to defects and environment. Greatly suppressed unintentional doping and large charge neutrality point shift equivalent to $3\times10^{12}$ electrons $cm^{-2}$ has been realized in monolayer $MoS_2$ and $WS_2$ samples on CYTOP substrates. Subsequently, ambipolar transport behavior, both positive and negative trions, have been observed in exfoliated monolayer $WS_2$ sample on CYTOP for the first time by using silicon back gate. The role of bandgap renormalization, Pauli blocking and carrier screening has been studied in a relatively clean environment with minimum substrate-induced doping and dielectric screening. The interplays between these many-body effects determine the blue-shift and red-shift of optical bandgaps in monolayer $MoS_2$ and $WS_2$, respectively, with the increase of carrier doping. Accessing the p-doped region by simple electrical control in TMDCs is also ideal for investigating novel valley contrasting phenomena, such as valley hall effect and valley pseudospin[22-24]. Furthermore, the CYTOP fluoropolymer can be uniformly coated onto large area and flexible surfaces, thus offering the capability to engineer the optoelectronic properties of 2D materials at wafer scale.

## Experimental Section

*Exfoliation and transfer of MX$_2$ flakes:* CYTOP was purchased from Asahi glass Co., Ltd. The CYTOP solution is diluted with solvent by 1:9 ratio and spin coated at 2000 rpm on SiO$_2$/Si to get a uniform film with thickness of 20 nm. The film was annealed at 100°C for 30 minutes and then 150°C for 1 hour to completely dry the film. Graphite, MoS$_2$ and WS$_2$ bulk crystals were purchased from Graphene-supermarket, SPI and HQ Graphene, respectively. Monolayer crystals were mechanically exfoliated on silicon substrates covered by a 300 nm layer of thermal oxide or on PDMS stamp. PDMS stamp has the advantages to obtain larger size monolayer flakes and reduce the amount of glue on the substrates. PDMS were soaked in acetone for hours to deplete the low molecular weight moieties before use. Monolayer crystals on PDMS were slowly transferred on to target substrates such as quartz, SiO$_2$/Si and CYTOP coated SiO$_2$/Si or quartz with a micro-manipulator under microscope. The PDMS is then slowly released. Mild heating is required to improve the adhesion when transferring onto CYTOP coated substrates. The monolayer samples were cleaned with acetone/ IPA after the transfer. We have compared the optical and electrical properties of the monolayer sample prepared by direct exfoliation on SiO$_2$/Si substrates and by exfoliation on PDMS then transferred onto SiO$_2$/Si, and samples produced by both methods showed practically no difference. Direct exfoliation on CYTOP coated substrates has very low yield and thus we used PDMS exfoliation and transfer method for all samples on CYTOP.

*Device fabrication:* We used micro stencil mask to make electrical contacts on the monolayer crystals. It is beneficial that the samples do not touch PMMA or other resist polymers, which are known to leave residues on the samples. The fabrication details of micro stencil mask are described in the Supporting Information Note 4 and Figure S4. The fabrication of FET devices is discussed in Figure S5, S6 and S7.

*Optical measurements:* All optical spectra were collected using micro-Raman system in a back scattering geometry and low temperature measurements are done with samples in a liquid nitrogen-flow cryostat. Photoluminescence spectra were obtained with a 532 nm excitation laser at intensities less than 20 μW. A tungsten-halogen lamp was used as a light source for the differential reflectance measurements. The light illuminating the sample was focused down to a ~2 μm spot using a small confocal pinhole. The intensity of the light on the sample was kept low in order to avoid sample damage or heating effect.

## Figure captions

**Figure 1. (a)** PL spectra of monolayer $MoS_2$ on various substrates at room temperature. **(b)** Optical image of a transferred monolayer $MoS_2$ flake on 20 nm CYTOP film on $SiO_2/Si$ substrate. The scale bar is 20 µm. Inset: AFM image of selected area in (b) with the height profile along the dashed line.

**Figure 2. (a)** PL spectra (cyan) and differential reflectance spectra of monolayer $MoS_2$ (blue) and monolayer $WS_2$ (red) on quartz substrate measured at room temperature. The Stokes shift is 25 and 15 meV for monolayer $MoS_2$ and $WS_2$, respectively. Inset: a schematic diagram showing the origin of Stokes shift due to the spatial fluctuation of exciton/trion states caused by substrate-induced inhomogeneous doping. **(b)** PL spectra (orange) and differential reflectance spectra of monolayer $MoS_2$ (blue) and monolayer $WS_2$ (red) on CYTOP/quartz substrate. The Stokes shift is negligible for both $MoS_2$ and $WS_2$. **(c)** Transfer curves of typical $MoS_2$ FET devices on CYTOP/$SiO_2$/Si (orange) and $SiO_2$/Si (cyan) substrates. The threshold voltage is -7 and -40 V, respectively. Inset: a schematic of a back-gated TMDC FET device. **(d)** Transfer curves of typical $WS_2$ FET devices on CYTOP/$SiO_2$/Si (orange) and $SiO_2$/Si (cyan) substrates. The threshold voltage for (n-type) electron transport is 10 and -50 V, respectively. The threshold voltage for (p-type) hole transport on CYTOP is -40 V.

**Figure 3. (a)** PL intensity of monolayer $MoS_2$ on CYTOP (orange) and $SiO_2$ (cyan) substrates as a function of $V_g$ at room temperature. Inset: Selected PL spectra of the monolayer $MoS_2$ on CYTOP/$SiO_2$/Si substrate at various $V_g$. The PL intensity is around 200 times stronger at -40 V $V_g$ than that at 80 V $V_g$. **(b)** The 2D map of differential reflectance of monolayer $MoS_2$ as a function of $V_g$ measured at 77K, showing the evolution of neutral exciton (A) and negative trion (A-) states. **(c)** Peak position of excitons (orange circles) and trions (red circles) extracted from (b). The calculated changing rate of $\Delta E_{PB}$ (cyan triangles), $\Delta E_b$ (blue stars) from Ref. 44 and $\Delta E_g$ (olive squares) as a function of carrier density, offset at the peak position of A exciton at charge neutral point. The charge neutral point here is estimated to be -40 V. **(d)** A schematic diagram of the evolutions of quarsiparticle ($E_g$) and optical ($E_{opt}$) bandgaps and exciton binding energy with electron doping. The grey dashed curves indicate the dark states arising from conduction band splitting at K/K' points caused by spin-orbital coupling in monolayer $MoS_2$. Left: no electron doping; right: at relative low doping density. The horizontal dotted line is a guide for the eyes.

**Figure 4. (a)** Differential reflectance spectra (red curves) with fittings (dark grey, cyan and orange curves) of monolayer $WS_2$ in the range of 1.9–2.1 eV with various back-gate voltages measured at 77K. The charge neutral point for this monolayer $WS_2$ sample is estimated to be -5 V. **(b)** The evolution of peak positions of the A exciton, negative trion (A-) and positive trion(A+). Two insets

depict the schematic illustrations of positive trion (left) and negative trion (right) involving the defect states. The grey short-dashed lines associated with A excitons are guides for the eyes. **(c)** A schematic diagram of the evolutions of quarsiparticle ($E_g$) and optical ($E_{opt}$) bandgaps and exciton binding energy with electron doping. The grey dashed curves indicate the dark states arising from valence band splitting at K/K' points caused by spin-orbital coupling in monolayer $WS_2$. Left: no electron doping; right: at relative low doping density. The horizontal dotted line is a guide for the eyes.

**Supporting Information**

Supporting Information is available from the Wiley Online Library or from the author.

**Acknowledgements**

B.L. and W.Z. contributed equally. We thank Dr. Kok Wai Chan (Centre for Advanced 2D Materials and Graphene Research Centre, NUS) for his helpful discussions. The authors acknowledge the National Research Foundation, Prime Minister Office Singapore, under its Medium Sized Centre Program.

**Figure 1**

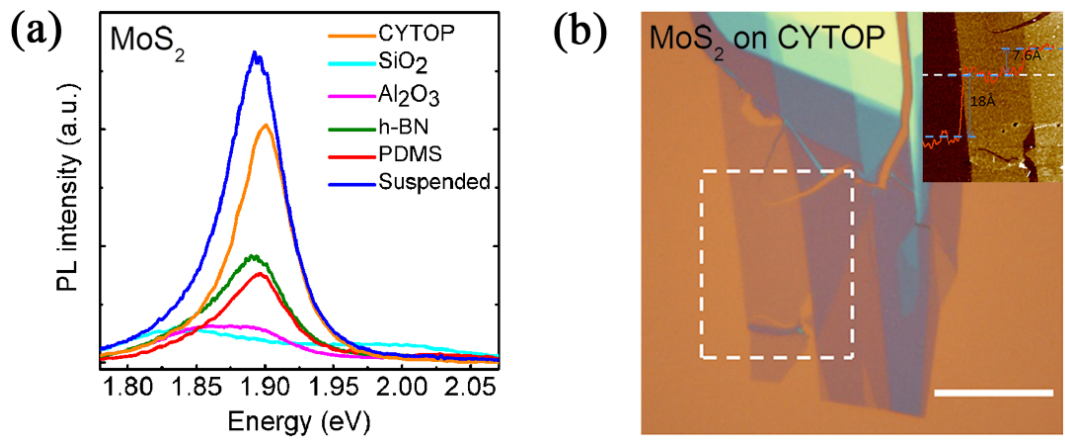

**Figure 2**

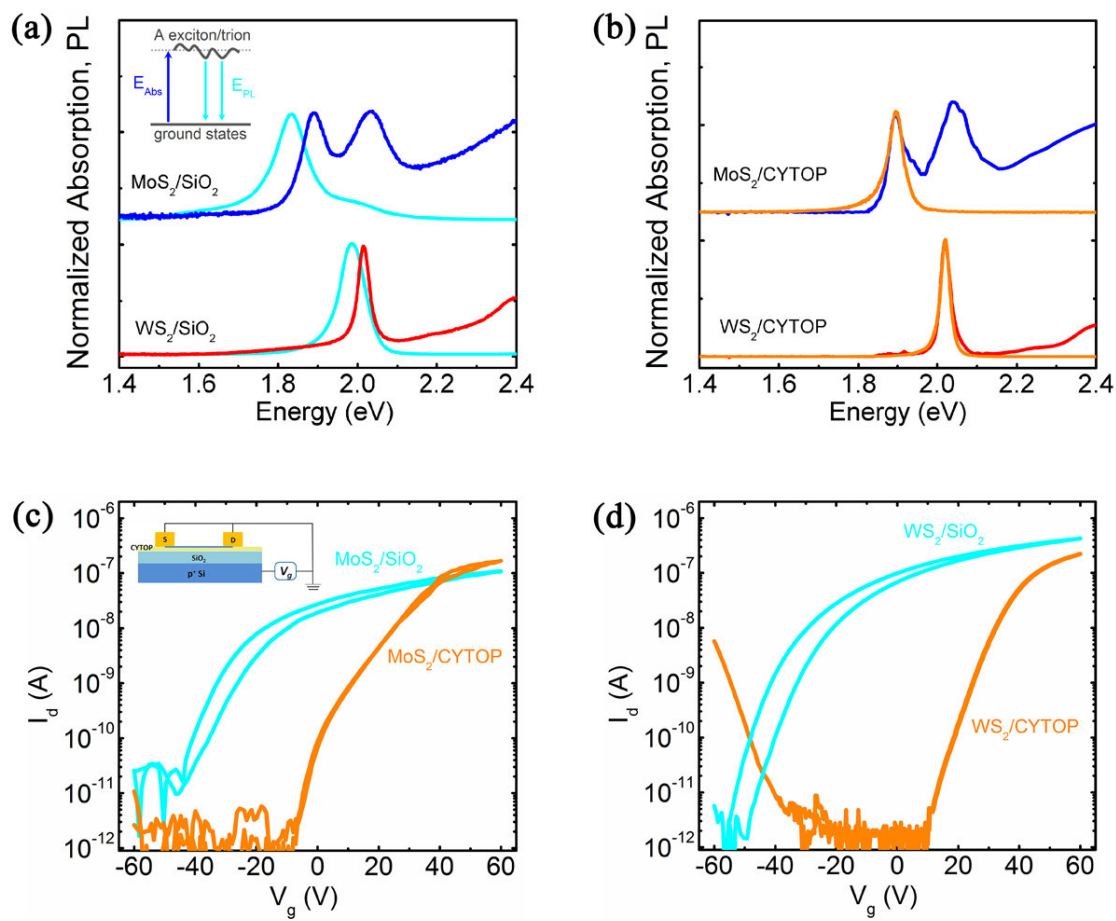

**Figure 3**

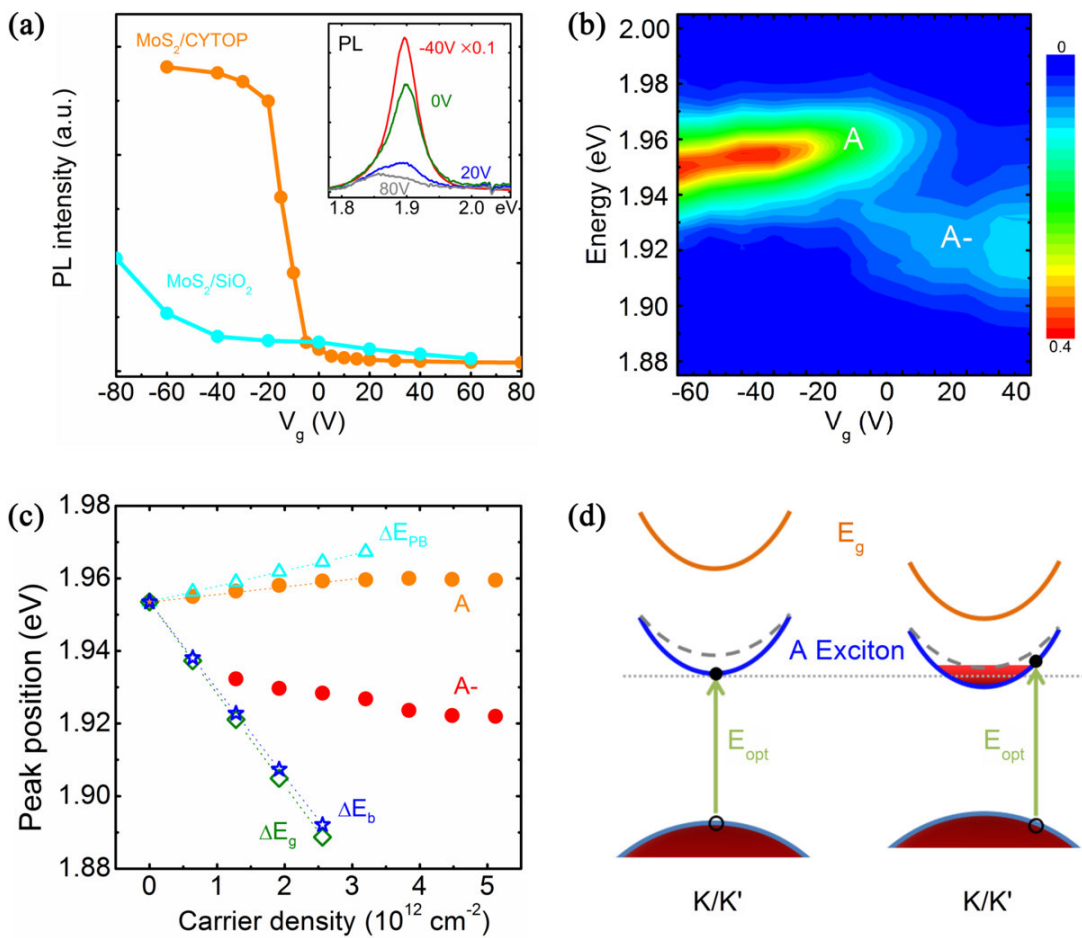

**Figure 4**

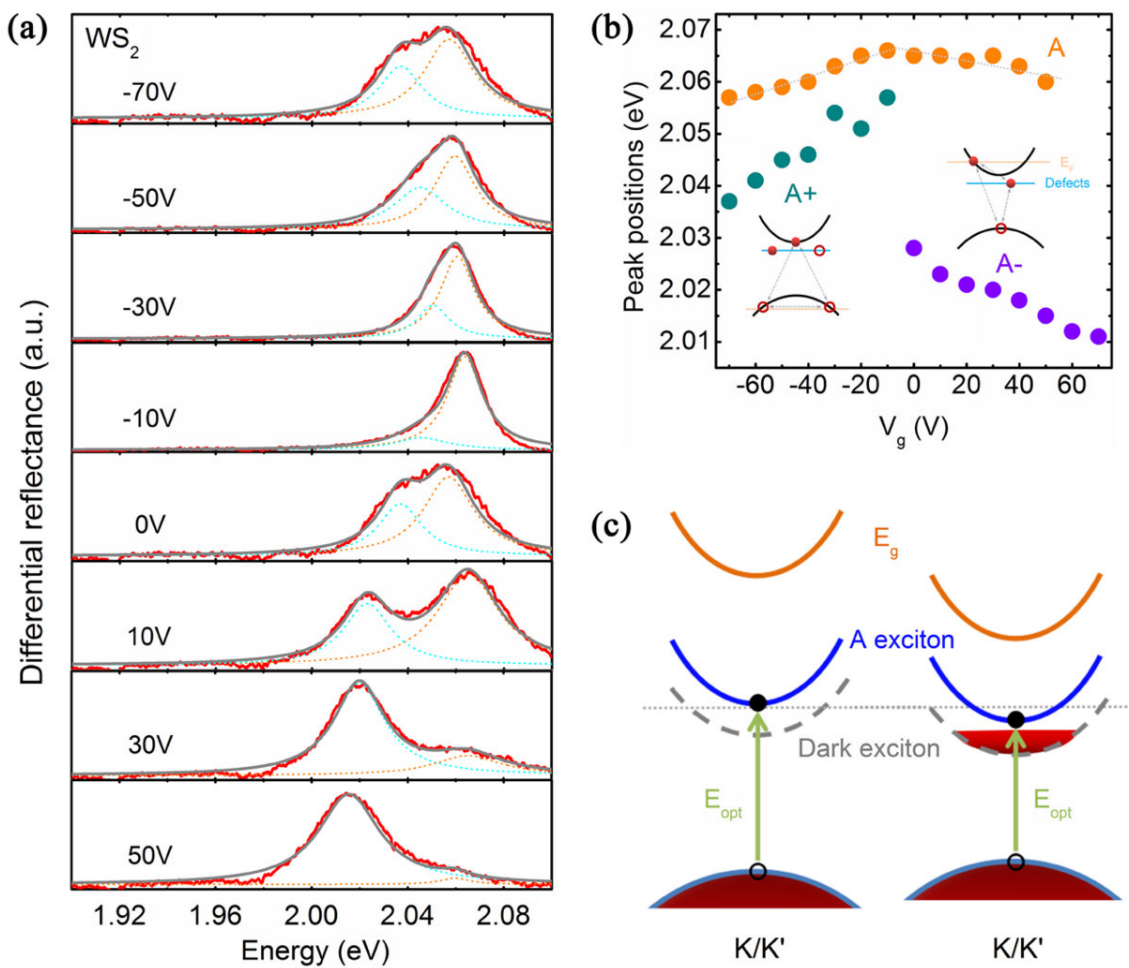